# HEPCloud, an Elastic Hybrid HEP Facility using an Intelligent Decision Support System

*Parag Mhashilkar*[1,*], *Mine Altunay*[1], *Eileen Berman*[1], *David Dagenhart*[1], *Stuart Fuess*[1], *Burt Holzman*[1], *James Kowalkowski*[1], *Dmitry Litvintsev*[1], *Qiming Lu*[1], *Alexander Moibenko*[1], *Marc Paterno*[1], *Panagiotis Spentzouris*[1], *Steven Timm*[1], *Anthony Tiradani*[1], *Eric Vaandering*[1], *John Hover*[2], and *Jose Caballero Bejar*[2]

[1] Fermi National Accelerator Laboratory, Batavia, Illinois, USA
[2] Brookhaven National Laboratory, Upton, New York, USA

**Abstract.** HEPCloud is rapidly becoming the primary system for provisioning compute resources for all Fermilab-affiliated experiments. In order to reliably meet the peak demands of the next generation of High Energy Physics experiments, Fermilab must plan to elastically expand its computational capabilities to cover the forecasted need. Commercial cloud and allocation-based High Performance Computing (HPC) resources both have explicit and implicit costs that must be considered when deciding when to provision these resources, and at which scale. In order to support such provisioning in a manner consistent with organizational business rules and budget constraints, we have developed a modular intelligent decision support system (IDSS) to aid in the automatic provisioning of resources spanning multiple cloud providers, multiple HPC centers, and grid computing federations. In this paper, we discuss the goals and architecture of the HEPCloud Facility, the architecture of the IDSS, and our early experience in using the IDSS for automated facility expansion both at Fermi and Brookhaven National Laboratory.

## 1 Introduction

HEPCloud is becoming scientific gateway to computing resources for all Fermilab-affiliated experiments and is rapidly becoming the primary system for provisioning the resources for all Fermilab-affiliated experiments. This provisioning is responsible for managing time allocations and monetary budget usage. It spans facilities including the High Performance Computing (HPC) centers such as Cori at the National Energy Research Scientific Computing Center (NERSC) and commercial clouds like Google Compute Engine (GCE) and Amazon Web Services (AWS). As part of the Fermilab HEPCloud project [1], we have been constructing an intelligent decision support system (IDSS) [2]. Our IDSS, known as the Decision Engine [3][4], provides the automation of requests for computing resource allocations across all participating experiments and affiliated facilities. An overall goal of the Decision Engine (DE) is to use both administration-defined and management-defined policies to create resource scheduling requests on behalf of the HEPCloud facility. The DE is responsible for ensuring that policies are applied in a reliable, traceable and consistent manner. The policies that are carried out ultimately result in resource requests, and ensure that these requests match incoming job requirements. Included in the DE is a software framework with stages for acquiring data, performing data analytics, and generating decisions using an inference engine. A knowledge base is used to manage all data made available within the running system. Careful attention is paid to the system-wide configuration coherency, addressing the needs of all user groups.

[*] Corresponding author: parag@fnal.gov

In this paper we describe the goals and high level architecture of the HEPCloud facility, architecture of the DE and our early experience in using the DE for automated facility expansion at Fermi and Brookhaven National Laboratory.

## 2 The HEPCloud Facility

The Fermilab scientific computing staff supplies software and services to support the physics program and provide essential resources for leading high energy physics (HEP) experiments including US-CMS, NOvA, g-2, and MicroBooNE, along with future experiments DUNE and mu2e. These resources include several types of dedicated and shared resources (CPU, disk, hierarchical storage, including disk cache, tape, tape libraries), for both data intensive and compute intensive scientific work. Support for these resources is currently limited to resources provisioned by and hosted at Fermilab, or to remote resources made available through the Open Science Grid (OSG) [5]. Expanding beyond the traditional HEP computing grid is essential to supplying the ever-growing need for simulation, experimental data processing, and end-user analysis.

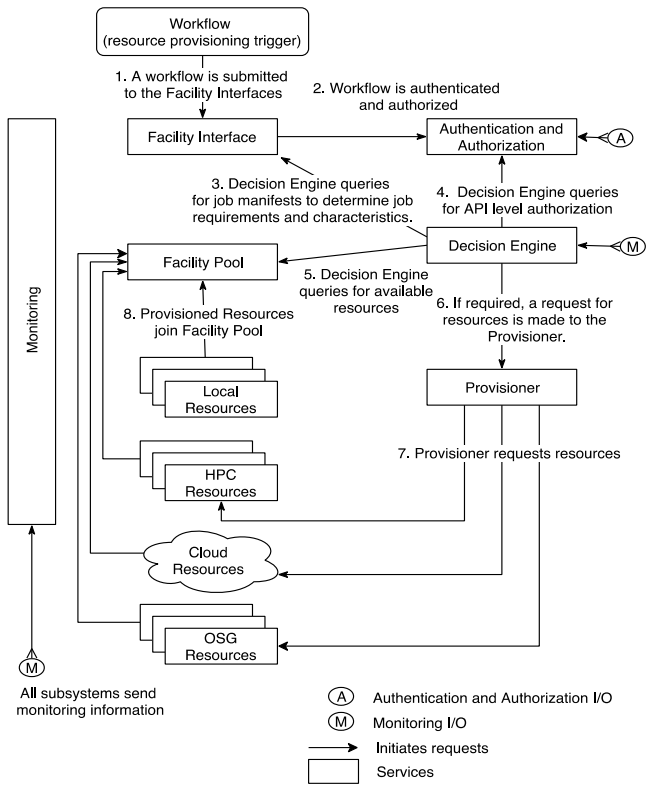

**Fig. 1.** Components involved in provisioning resources in the HEPCloud Facility.

In order to reliably meet peak demand, Fermilab must plan to provision enough resources to cover this forecasted peak, rather than using some other statistic such as median demand. This can be cost ineffective, since some resources may be underutilized during non-peak periods even when resource sharing (enabled by HEP grids) is accounted for. Scientific productivity will be affected if the forecasted demand is too low, since there is a long lead time to significantly increase the use of local or remote resources. HEPCloud intends to mitigate these problems by intelligently extending the current Fermilab compute facility to execute jobs submitted by scientists on a diverse set of resources, including commercial and community clouds, grid federations, and HPC centers. This will allow provisioning in a more efficient and cost-effective and elastic way. Additionally, this will enable the facility to respond to demand peaks without over-provisioning local resources.

Figure 1 shows various components in the HEPCloud facility. It also shows a typical flow of events involved in provisioning compute resources to run jobs in the facility. The DE periodically queries different systems and services to identify computational jobs in various job queues that are in need of compute resources. Based on the jobs and resource manifests, the DE short lists candidate resources eligible to run these jobs. The DE uses the price-performance metrics of these resources, their costing information, current occupancy and state. It then applies administrative-defined and management-defined policies to generate resources requests that are used by a provisioner to expand the facility.

## 3 Decision Engine

New capabilities are continuously being made available for the open market as commercial clouds continue their fast-pace adoption and deployment of new computing technologies. At the same time, HPC sites are beginning to target scientific communities that run applications that differ from the MPI-style fine-grained parallel jobs common to these machines. These new communities are influencing the design of future HPC machines. This presents us with an opportunity to make use of resources at scales which were previously unreachable. These new capabilities, along with the desire to satisfy the ever-increasing demand for compute resources, has led to new and existing experimental communities to request access to these machines and providers. As a result, a new way of matching job requirements to resource capabilities and associated costs is now necessary.

Both commercial cloud and allocation-based HPC resources have explicit and implicit costs that must be considered during resource provisioning. Commercial clouds require payment in monetary currencies, while HPC sites grant allocations in wall hours. Grid federations such as the OSG offer their resources on an opportunistic basis. The DE must use algorithms to compare relative costs to relative value of running jobs to be able to determine the best mix of resources to satisfy customer and management needs. The DE must compare the requirements of the workflow with the requirements of the facility, which may include budgetary constraints, while enforcing administrative and management policies.

### 3.1. Decision Engine Architecture

The primary drivers of the DE design were: (1) the need for a framework that enforces the processing stages defined and implemented by the program, and which provides for the injection of user-supplied code and expert knowledge; (2) the need for a configuration and assembly system that instantiates the appropriate user-supplied code, and that provides the necessary context-dependent information to realize different parameterizations of that code; and (3) a means to manage the data being processed and the varying timescales for the relevance and validity of those data.

Figure 2 shows the high-level architecture of the DE. The DE consists of a DataSpace acting as a Knowledge Management system, a ConfigManager acting as the Configuration Factory, and one or more TaskManagers responsible for the scheduling of processing. A single Engine is responsible for coordinating their work.

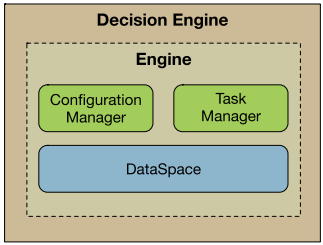
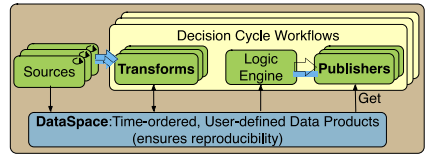

**Fig. 2.** The Decision Engine top-level components.    **Fig. 3.** Key elements on the design.

### 3.1.1 Decision Channel

The DE can manage and run algorithms of varying complexity for the purpose of requesting resources for computing jobs. The DE defines a Decision Channel as a grouping of tasks that generate a decision. A decision consists of a recommendation of one or more actions that should be executed (such as allocation of computing resources), actions that are directly executed (such as updating of monitoring systems), or both. The modularity provided by Decision Channels allows the DE to manage decision making processes as distinct units. A Decision Channel can be brought online or offline as needed, thus changing the state of the entire DE. Because of the modular design, this feature allows different algorithms to be developed and tested independently by different domain experts.

Each Decision Channel task, implemented as a Python class, contains several modules, each of which adheres to a common protocol and interface. We currently define four module types: Source, Transform, Logic, and Publisher. A Decision Channel minimally consists of one of each of these kinds of modules as shown in Figure 3. Each module adheres to a specific contract that governs how the modules connect. For example, each module (except Sources) expresses the names of all the data products the module consumes, and (except for Publishers) the names of all the data products the module produces.

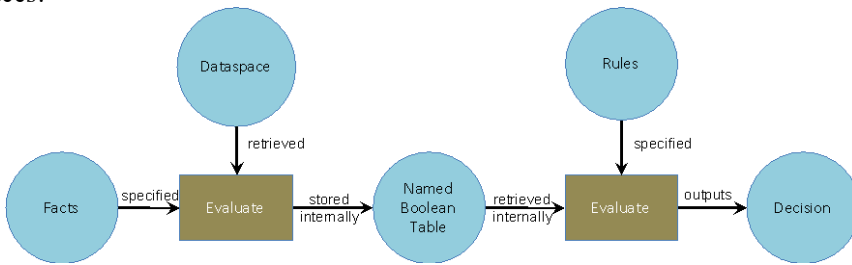

**Fig. 4.** Logic engine: Evaluation of facts and rules.

Each Source module is scheduled periodically by the framework. A Source is responsible for communicating with an external system (via the native APIs of that system) to gather data that acts as input to the decision-making process. A Transform module contains algorithms to convert input data into new data. A Transform consumes one or more data products (produced by one or more Sources, Transforms, or both) within a Decision Channel, and produces one or more new data products. A runtime error is produced if one of the input data products does not exist. Transforms are expected to produce the data products they claim they will produce according to their contract. The Logic Engine (Figure 4) is a rule-based forward-chaining inference based on the Message Analyzer package [6], a software product developed and supported in-house.

The Logic Engine operates on "facts". Each fact has a name and an expression that evaluates to a boolean. The value of a fact is the value of the expression. Expressions can access and operate on data produced by Source and Transform modules. A rule consists of a condition composed of references to facts and boolean operations on their values. Actions are triggered when the rule evaluates to boolean "True". Logic Engine rules can produce new facts that evaluate to the result of the rule's boolean expression. This fact can be used by subsequent rules. In this manner, rules can be developed separately as blocks and chained together. The Logic Engine processes all facts, followed by all rules to obtain the final decision. When processing rules, there is no inherent ordering implied. A Publisher module is the inverse of a Source. Publishers consume data products produced by Sources and Transforms. They use remotely exposed APIs to push the data products to the external systems. In a complex system defining several Decision Channels it is quite possible for data acquired by a Source to be usable by transforms from more than one Decision Channel. A special type of Source called Source Proxy is used to retrieve data from a different Decision Channel.

### 3.1.2 Knowledge Management System

The DataSpace is the knowledge management system of the DE. It is a time-sensitive data store that contains the complete state of the DE. All data management, archiving, and time management is accomplished through the management of self-contained private "databases" called DataBlocks. These DataBlocks are assigned to Decision Channels and contain all data gathered by Sources, all data generated by Transforms, all results from Logic Engines, and all data required for traceability, debugging, and logging. The DataSpace manages the versioning and archiving of DataBlocks.

### 3.1.3 Decision Cycle

A Decision Cycle is a scheduling concept that is implemented in the context of a Decision Channel. When a Decision Channel starts, it enters an initialization phase. All Source modules must run at least once before the channel is considered to be in steady state operation. This is considered the operational or running state. Sources always put generated data in the DataBlock versioned as *tnext*, where *tnext* represents the state of the system to be used by the next Decision Cycle. While in the operational state, a data product added to the DataBlock from any Source module triggers a Decision Cycle. When a Decision Cycle starts, the complete DataBlock is copied and versioned as *tcurr*. All interactions with the DataBlock during the Decision Cycle occur on the most recent backed up version, or *tcurr* while the running sources continue to add dataproducts to *tnext*.

The first stage of the Decision Cycle is to run all configured Transforms followed by the Logic Engine. The Logic Engine then processes all facts and applies rules to produce a result that consists of a fact containing the publishers that should be run. This fact is stored in the *tcurr* version of the DataBlock. The publishers named by the Logic Engine then use the data products in the DataBlock to send information to external systems. At the end of the Decision Cycle the DataBlock *tcurr* is locked and permanently archived. The system does not allow a particular Decision Channel to run multiple Decision Cycles at any given time. This is enforced by preventing sources from triggering a Decision Cycle when a Decision Cycle is already in progress.

### 3.1.4 Task Manager

The Task Manager is the implementer of a configured Decision Channel. It is responsible for scheduling and executing the configured modules. At any given time, a DE can be configured to run several asynchronous Decision Channels in parallel. Separating the runtime environment of the Decision Channels from the core framework component increases robustness by shielding the system from badly implemented plugins or badly configured Decision Channels. This separation also allows the DE to manage different task managers independently. The DE can safely shutdown a task manager executing a Decision Channel without impacting other Task Managers.

## 4 Experience with DE

### 4.1 Decision Engine with glideinWMS as the Resource Provisioner

We performed an integration test to demonstrate the functioning of the system, and to evaluate the effectiveness of our design choices in a simplified but realistic setting. Our evaluation of the prototype investigated three main aspects of the DE: the separation of roles and responsibilities of the algorithm developers from the operators of the DE service, the expression of business requirements using facts and rules that can be interpreted by the inference engine, and the configuration of multiple Decision Channels that can be aggregated by the DE and the ability to execute them as per the instructions. We also used our evaluation to identify improvements we will make in our next round of development.

We modelled a scenario in which resources from AWS, NERSC, and opportunistic resources from the OSG were available to be provisioned by glideinWMS [7] and made part of the facility's local computing cluster. We introduced jobs that expressed a preference for one of the types of resource. We simulated a limited amount of funds for running on AWS and limited allocations at NERSC. Figure 5 shows resources provisioned during the DE evaluation.

The DE requested resources on three grid sites, one AWS availability zone and at NERSC. The DE periodically queried the HTCondor schedulers for idle jobs. The DE took into account preferences expressed by jobs and made provisioning requests to one or more available resource types. Provisioning requests were evenly distributed for jobs that preferred more than one resource type. During the evaluation over 1400 computation jobs for the CMS experiment were run on the provisioned resources.

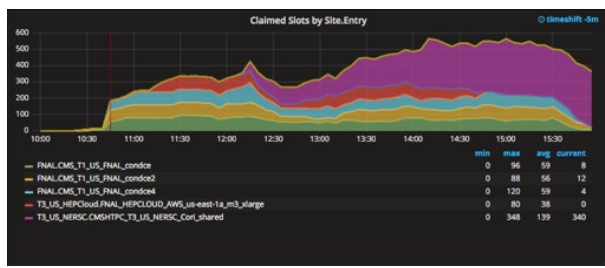

**Fig. 5.** Acquiring compute slots at AWS, NERSC Grid sites with the Decision Engine.

### 4.2 Decision Engine with VC3 as the Resource Provisioner

Brookhaven National Laboratory (BNL) is a participating institution in the HEPCloud project and plans to leverage the facility in the future. BNL developers are members of the core team for the Virtual Clusters for Community Computation project (VC3) [8]. VC3 is a web-based central service that allows groups of researchers to aggregate their resource allocations (e.g. at campus clusters and other computing centers) to run fully dynamic virtual clusters. These clusters are typically HTCondor batch but can also be Spark or JupyterHub clusters.

A typical use case would be an experiment with Principal Investigators and students at two universities, each of which have allocations on their local campus clusters. VC3 allows them to form a group project and pool their allocations, allowing both PIs and students to launch virtual clusters which span both campus clusters. The VC3 overlay abstracts away all heterogeneity so that each virtual cluster appears uniform, with the overlay satisfying dependencies as needed on each campus resource. Each virtual cluster is typically provided with a head node on the central infrastructure, where all project members may log in and work.

BNL developers wrote DE modules to instantiate a Decision Channel which requests resources via the VC3 system. This was implemented as a publisher that uses the VC3 client code to make virtual cluster requests. Rather than provisioning a head node on VC3 central resources, these virtual cluster workers are configured to connect back to the HEPCloud pool(s). The deployment and testing of the VC3-related functionality required only that a VC3 client credential (in this case a host certificate) be made available to the DE publisher. VC3 then trusts that publisher to make requests.

The primary benefit of this work was that it allowed a HEPCloud facility to provision on target resources that are not curated within HEPCloud. This avoids the need to go through the process of (re-)defining targets for HEPCloud that are already defined within VC3. In addition, due to the way VC3 instantiates virtual clusters, the DE can simply specify the target number and configuration of workers, and VC3 will satisfy the request as needed from the pool of target resources.

A secondary benefit of this work was that it demonstrated the ability of external domain experts (in this case VC3 developers at BNL) to easily and effectively write DE plugins to handle novel requirements. This interoperability validates the plugin/module system of the DE and demonstrates the flexibility of both project codebases.

## 5 Conclusion

The prototype supported only a subset of the features expected from the production version of the DE. Upcoming releases of the framework will provide full support for Source Proxies, essential for effectively connecting multiple Decision Channels. The DE will also include an improved Configuration Manager that will allow users to express policies. The Configuration Manager will be able to validate rules based on the available standard library, thus, minimizing chance of operator error or invalid rules.

We also plan to implement essential administrative tools to help the administrators to debug and troubleshoot Decision Channels, manage individual Decision Channels independently, perform configuration management via a user-friendly interface, provide support for high availability, etc. We also plan to perform more extensive testing of the DE to run at expected scales.

The current DE standard library implements functionality to generate resource provisioning requests based on the current status of the facilities pool. We plan to extend its functionality to incorporate network capacity and data locality into the decision-making process. For example, if jobs require heavy network I/O, requests can avoid resources with limited network connectivity. Also, the standard library can be expanded to factor in the impact of one or more potential decisions on AWS financial status and allocations at NERSC.

The DE design enforces the separation of the core framework from modules. Each module is a pluggable component. Its instantiation in the context of a Decision Channel can be customized using one or more configuration parameters. We found that this plugin-based approach makes it easier for separation of the responsibilities based on expertise between the module maintainers and the DE service operators. Since each module can also be easily invoked in a standalone manner, it makes it easier to perform unit and black box testing enabling larger teams to work in a disconnected fashion—provided the modules follow the data access protocol.

## Acknowledgment

This manuscript has been authored by Fermi Research Alliance, LLC under Contract No. DE-AC02-07CH11359 with the U.S. Department of Energy, Office of Science, Office of High Energy Physics. This research used resources of the National Energy Research Scientific Computing Center, a DOE Office of Science User Facility supported by the Office of Science of the U.S. Department of Energy under Contract No. DE-AC02-05CH11231.